# NEUTRON DIFFRACTION STUDIES ON $La_{2-x}Dy_xCa_{2x}Ba_2Cu_{4+2x}O_z$ SUPERCONDUCTORS

S. Rayaprol*, Rohini Parmar, D. G. Kuberkar, Keka R. Chakraborty+, P. S. R. Krishna+ and M. Ramanadham+

*Department of Physics, Saurashtra University, Rajkot – 360 005 (INDIA)*

+*SSPD, Bhabha Atomic Research Center, Mumbai – 400 085 (INDIA)*

## ABSTRACT

*Structural studies on Dy-substituted La-2125 type superconductors have been carried out by neutron diffraction experiments at room temperature using a monochromatic neutron beam of wavelength (λ) = 1.249Å. A series of samples with $La_{2-x}Dy_xCa_{2x}Ba_2Cu_{4+2x}O_z$ stoichiometric composition, for x = 0.1 – 0.5, have been studied for their structural properties. A tetragonal Y-123 unit cell was taken as the starting model for the Rietveld analysis. All the samples fit into the starting model, exhibiting no structural transition taking place with increasing dopant concentration. The results of Rietveld analysis and structural properties will be discussed in detail.*

## 1. INTRODUCTION

Superconductivity in $La_2Ba_2Cu_4O_z$ (La-224) can be induced by the simultaneous addition of CaO and CuO in the stoichiometric composition $La_{2-x}R_xCa_{2x}Ba_2Cu_{4+2x}O_z$ where R = rare earth, and x = 0.1 – 0.5 (i.e., La-2125 phase for x = 0.5) [1]. Several studies have been carried out in understanding the origin of superconductivity in these types of oxides [2-4]. The R ion in La-2125 type compounds provides structural stability without any effect on superconductivity similar to the role of R ion at Y-site in Y-123 superconductors. However, for R = Pr, we observed superconductivity even in the presence of higher concentrations of Pr, unlike in the case of Y-123 systems [5]. These interesting observations calls for detailed structural investigations on La-2125 type mixed oxide systems. In this paper we present the results of structural investigations on the R = Dy, substituted La-2125 type compounds studied by neutron diffraction experiment. The structural studies were carried out using the FULLPROF program [6]. In these compounds with increase in the dopant concentration from x = 0.1 the superconducting transition temperature ($T_c$) increases from 38 K to 75 K, the highest $T_c$ observed for x = 0.5 i.e., $La_{1.5}Dy_{0.5}Ca_1Ba_2Cu_5O_z$ compound.

## 2. EXPERIMENTAL DETAILS

All the samples in the $La_{2-x}Dy_xCa_{2x}Ba_2Cu_{4+2x}O_z$ for x = 0.1 – 0.5, (abbreviated as LDCBO) series were prepared by the solid-state reaction method. The details of the synthesis are reported in Ref 1. Superconducting transition temperatures were determined by dc four probe resistivity and dc susceptibility measurements. All the samples were examined for single-phase formation by X-ray diffraction method. Neutron powder diffraction data were recorded on samples of LDCBO series on a powder diffractometer ($\lambda$ = 1.249 Å) at Dhruva Reactor, BARC.

* Present Address: DCMP & MS, TIFR, Mumbai – 400 005 (INDIA) Email: sudhindra@tifr.res.in

## 3. RESULTS AND DISCUSSION

The structure of Ca doped tetragonal perovskite like R-224, R-1113, R-3137, and La-2125 has been has been found to be isostructural to the tetragonal R-123 phase with P4/mmm space group [7-10]. The normalization of La-224 and La-2125 phases can be written in the normalized La-123 form as

$$LaBa_2Cu_3O_{7-\delta} = (La_2Ba_2Cu_4O_z) \times (3/4) = La_{1.5}Ba_{1.5}Cu_3O_{z'} \text{ (where } z' = 3z/4)$$
$$= La_1(Ba_{1.5}La_{0.5})Cu_3O_{z'} \quad \text{--- (1)}$$

and,

$$LaBa_2Cu_3O_{7-\delta} = (La_2Ca_1Ba_2Cu_5O_z) \times (3/5) = La_{1.2}Ca_{0.6}Ba_{1.2}Cu_3O_{z'} \text{ (where } z' = 3z/5)$$
$$= (La_{0.6}Ca_{0.4})(Ba_{1.2}Ca_{0.2}La_{0.6})Cu_3O_{z'} \quad \text{--- (2)}$$

In our case $Dy^{3+}$ is substituted at $La^{3+}$ site, thus, for x = 0.5 in La-2125 form, we get

$$= (La_{0.3}Dy_{0.3}Ca_{0.4})(Ba_{1.2}Ca_{0.2}La_{0.6})Cu_3O_{z'} \quad \text{--- (3)}$$

in normalized (to La-123) form.

Factors (3/4) and (3/5) represent the ratio of number of copper ions in R-123 to the number of copper ions in La-224 and La-2125 formula respectively. Equations 1, 2 and 3 shows the normalized La-224, La-2125 and Dy substituted La-2125 phases respectively, which helps in understanding the distribution of various cations in the normalized form.

The structure of the normalized La-2125 unit cell is shown in Figure 1. In the La-2125 normalized form, given by equation (2), $Ca^{2+}$ is distributed at both $La^{3+}$ and $Ba^{2+}$ sites alongwith the concomitant displacement of $La^{3+}$ onto $Ba^{2+}$ site. Looking at the R-123 unit cell, we find two types of cations, 'A' (like $Y^{3+}$ or trivalent lanthanide cations, which have typical ionic radii ~ 0.9 – 1.3 Å for *coordination number* (*CN*) *6*) and 'B' type ($Cu^{2+}$ with ionic radii ~ 0.73 Å *CN 6*). $Ca^{2+}$ is thus 'A' type cation. Owing to the similar ionic radii of $La^{3+}$ and $Ca^{2+}$ (1.03Å and 1.00 Å, *CN 6* respectively), $Ca^{2+}$ is expected to prefer $La^{3+}$ than to $Ba^{2+}$ site (~ 1.35 Å *CN 6*) [11]. Interestingly, the structural analysis of the La-2125 type oxides confirms the presence of Ca at Ba site also. Though $Ca^{2+}$ at $La^{3+}$ is '*hole doping*', $La^{3+}$ at $Ba^{2+}$ is '*hole filling*' it has been found that the simultaneous substitution of Ca at La and La at Ba is non-compensatory which can be seen by the increase in hole concentration and $T_c$ [3, 8].

According to the normalization shown in equations 1-3, a starting model of tetragonal La-123 unit cell was assumed to begin the structural analysis for all LDCBO samples. Figure 2 shows a typical neutron diffraction pattern for x = 0.1 sample of the LDCBO series. The theoretically calculated pattern, which has been refined into the observed data, can also be seen as the continuous line passing through the observed data points.

During refinement the structural parameters like unit cell constants (a, b, c), half width parameters (U, V, W) were refined in the first step. Secondly, atomic positions (only Z in this case), thermal parameters (B) were refined. Finally, site occupancies were refined to get a good agreement between calculated and observed patterns. Refinement for all the patterns converged with reliable Rietveld profile factors (***R-factors***). Table 1 lists in detail various crystallographic parameters derived from the Rietveld analysis of neutron diffraction data for all samples studied.

The variation in the unit cell parameters (a, c and Volume) with increasing dopant concentration is plotted in Figure 3(a-c). With the increase in proportion of smaller ionic radii i.e., $Dy^{3+}$, $Ca^{2+}$ at $La^{3+}$ and $La^{3+}$, $Ca^{2+}$ at $Ba^{2+}$ sites respectively, there is shrinking of the unit cell, which is evident from the decrease in unit cell constants. There is a significant decrease in the unit cell parameters from x = 0.1 to 0.3, but for x = 0.3 to 0.5, the unit cell parameters are almost same as if it has reached saturation value. It is interesting to note here that, with the decrease in the cell constants, we observe increase in the superconducting transition temperature, which is shown in Fig 3 d. Figure 3, shows that the changes in the values of unit cell parameters does effect the superconducting transition temperature for LDCBO samples.

The total percentage occupation of 1*d* site (rare earth site) by $Ca^{2+}$ increases significantly from 14 % to 42 % as the dopant concentration is increased from x = 0.1 to 0.5 (Table 2). Similarly, the simultaneous displacement of $La^{3+}$ to 2*h* (Ba) site increases marginally from 26% to 32%. The percentage of $Ca^{2+}$ at 2*h* site remains around 12 – 15. Thus, increase in superconducting transition with increasing dopant concentration can be attributed to the increase in the $Ca^{2+}$ concentration, which in turn increases the hole concentration per unit cell. During refinement of the neutron data, the rare earth ($Dy^{3+}$) was kept fully occupied at 1*d* site. Owing to the similarity of ionic radii of $La^{3+}$ and $Ba^{2+}$, we see concomitant displacement of $La^{3+}$ onto $Ba^{2+}$ site. Refinement cycles with distribution of $Dy^{3+}$ at both $La^{3+}$ and $Ba^{2+}$ site did not yield reliable ***R-values***. Hence, $Dy^{3+}$ was kept full at R site only (i.e., at 1*d*) during final refinement where the refinement converged with reliable ***R-values***. Thus, the role of $Dy^{3+}$ in these compounds is to stabilize the structure. This is in tune with the observation of role of rare earths in the R-123 structures, where the moment of the R (except $Ce^{3+}$, $Pr^{3+}$ and $Tb^{3+}$) has no effect on the superconducting properties. The small R substitution has been found to stabilize the La-2125 phase, since pure La-2125 phase is difficult to form.

## 4. CONCLUSIONS

The increase in dopant concentration ($Ca^{2+}$ and $Dy^{3+}$) results in increase of superconducting transition temperature for LDBCO compounds. Role of $Dy^{3+}$ has been found to stabilize the crystal structure, and has no pronounced effect on the superconducting transition temperature. $Ca^{2+}$ substitution helps in inducing superconductivity by creating holes at $La^{3+}$ site resulting in, presumably, bridging of two conductive $CuO_2$ sheets. The Rietveld analysis of the neutron diffraction data shows a systematic increase of $Ca^{2+}$ at $La^{3+}$ site with concomitant displacement of $La^{3+}$ onto $Ba^{2+}$ site, thus presenting a situation of both 'hole doping' and 'hole filling'. The increase in superconducting transition temperature thus confirms the fact that both these mechanisms are 'non-compensatory' and 'hole doping' dominates over 'hole filling' resulting in the induction of superconductivity.

Finally, $Ca^{2+}$ plays a significant role in introducing superconductivity in $La_{2-x}Dy_xCa_{2x}Ba_2Cu_{4+2x}O_z$ system without inducing any structural transition.

## ACKNOWLEDGEMENTS

The present work was carried out under the IUC-DAEF project number CRS-M-88 of Dr. D. G. Kuberkar with encouraging support from the SSPD group of BARC (India). SR and RP are thankful to IUC-DAEF, Mumbai for financial support.

Table 1 Values obtained from Rietveld analysis of Neutron diffraction data

| ***Concentration 'x' →*** | **0.1** | **0.2** | **0.3** | **0.4** | **0.5** |
|---|---|---|---|---|---|
| $N_{La}$ ***(1d)*** | 0.809 (2) | 0.614 (3) | 0.591 (3) | 0.483 (2) | 0.319 (4) |
| $N_{Dy}$ | 0.046 (2) | 0.136 (3) | 0.191 (3) | 0.233 (2) | 0.319 (4) |
| $N_{Ca}$ | 0.146 (2) | 0.250 (3) | 0.19 1 (3) | 0.233 (2) | 0.419 (4) |
| $N_{Ba}$ ***(2h)*** | 1.455 (5) | 1.345 (9) | 1.366 (7) | 1.317 (7) | 1.253 (10) |
| $N_{La}$ | 0.548 (5) | 0.595 (9) | 0.566 (7) | 0.567 (7) | 0.653 (10) |
| $N_{Ca}$ | 0.053 (5) | 0.004 (9) | 0.266 (7) | 0.317 (7) | 0.253 (10) |
| ***Z*** | 0.183 (32) | 0.184 (59) | 0.186 (40) | 0.186 (37) | 0.187 (52) |
| $N_{Cu1}$ ***(1a)*** | 1.000 | 1.000 | 1.000 | 1.000 | 1.000 |
| $N_{Cu2}$ ***(2g)*** | 2.0000 | 2.0000 | 2.0000 | 2.0000 | 2.0000 |
| ***Z*** | 0.348 (21) | 0.352 (40) | 0.353 (29) | 0.353 (27) | 0.355 (38) |
| $N_{O1}$ ***(2f)*** | 0.836 (24) | 0.848 (41) | 0.722 (30) | 0.720 (28) | 0.695 (42) |
| $N_{O2}$ ***(2g)*** | 2.286 (32) | 2.197 (53) | 2.423 (41) | 2.423 (37) | 2.364 (54) |
| ***Z*** | 0.158 (44) | 0.159 (78) | 0.160 (51) | 0.160 (47) | 0.162 (66) |
| $N_{O4}$ ***(4i)*** | 3.941 (26) | 3.774 (45) | 3.910 (32) | 3.910 (29) | 3.905 (42) |
| ***Z*** | 0.365 (20) | 0.367 (40) | 0.368 (26) | 0.368 (23) | 0.369 (34) |
| ***a = b (Å)*** | 3.895 (4) | 3.881 (4) | 3.860 (4) | 3.860 (4) | 3.861 (4) |
| ***c (Å)*** | 11.713 (3) | 11.682 (3) | 11.635 (3) | 11.635 (3) | 11.635 (3) |
| ***Volume*** $(Å)^3$ | 177.79 (2) | 176.04 (3) | 173.38 (2) | 173.38 (2) | 173.4 (2) |
| ***Total Oxygen*** | | | | | |
| ***(z' – in 123)*** | 7.06 | 6.82 | 7.05 | 7.05 | 6.96 |
| ***(z – in 2125)*** | 9.88 | 10.00 | 10.81 | 11.28 | 11.60 |
| ***R-factors*** | | | | | |
| $\chi^2$ | 1.98 | 4.16 | 2.25 | 1.89 | 2.99 |
| $R_{wp}$ | 6.51 | 9.90 | 6.98 | 6.39 | 7.77 |
| $R_{exp}$ | 4.63 | 4.85 | 4.65 | 4.65 | 4.49 |
| ***Bragg-R*** | 7.77 | 9.20 | 12.1 | 8.81 | 14.3 |
| $R_f$ ***– factor*** | 6.25 | 7.84 | 10.2 | 6.47 | 11.2 |
| ***Fractional (%)*** | 100.00 (0.92) | 100.00 (1.61) | 100.00 (1.28) | 100.00 (1.18) | 100.00 (1.74) |

Oxygen content per unit cell = z'; Oxygen content per formula unit = z

Figure Captions

Figure 1 The structure of the La-2125 unit cell, derived from the tetragonal La-123 structure.

Figure 2 A typical neutron diffraction profile of the $La_{2-x}Dy_xCa_{2x}Ba_2Cu_{4+2x}O_z$, x = 0.1 sample. The observed data points are shown as open circles, and the pattern generated after Rietveld structure refinement is plotted as a continuous line. The agreement of the observed and the calculated profile can be clearly seen in the insert, which shows the data between (2θ) $50^0$ – $70^0$ degrees.

Figure 3(a-d) The variation of unit cell parameters a, c, Volume and superconducting transition temperature ($T_c^{on}$) are plotted as a function of dopant concentration (x).

Figure 1

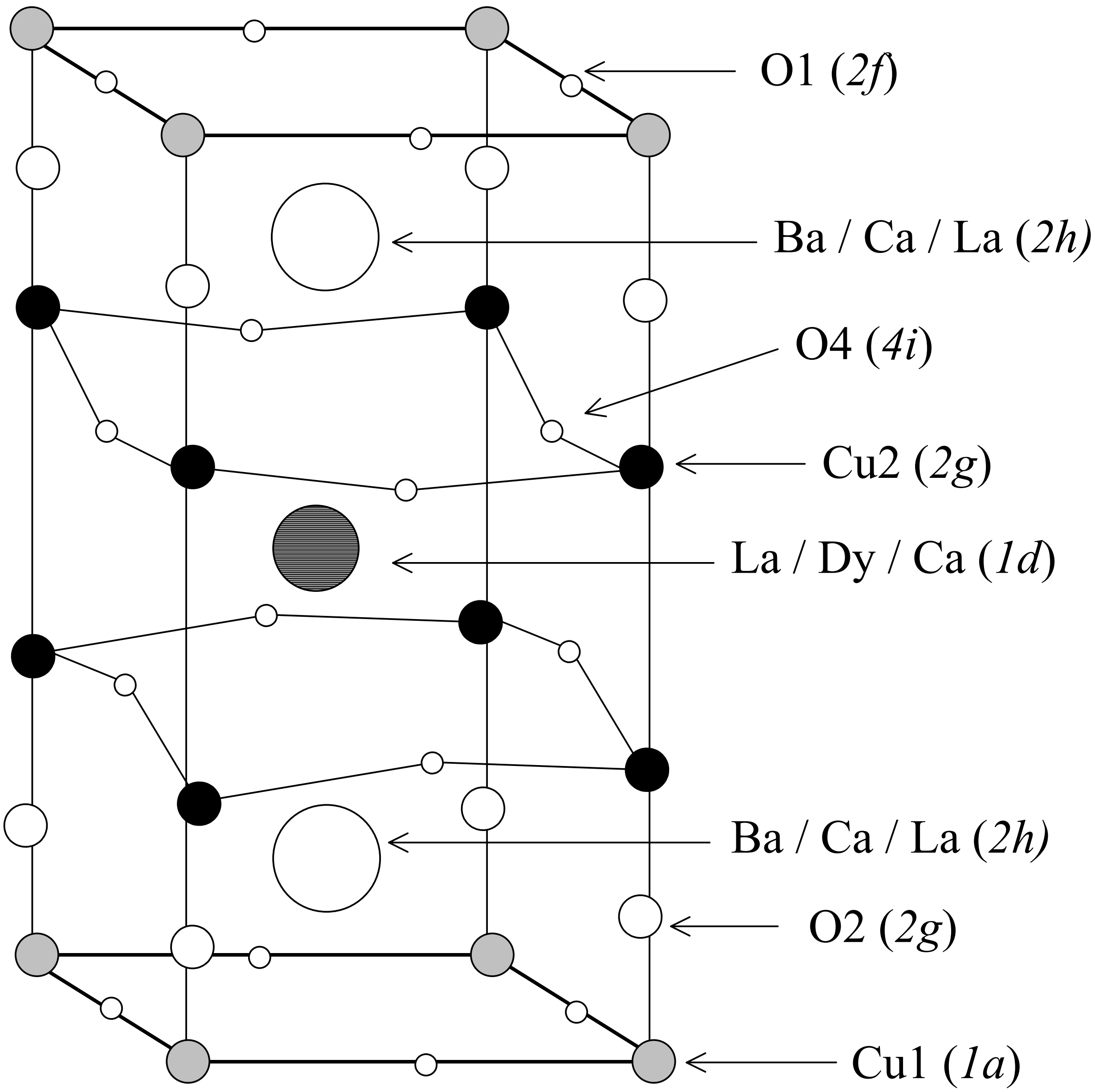

Figure 2

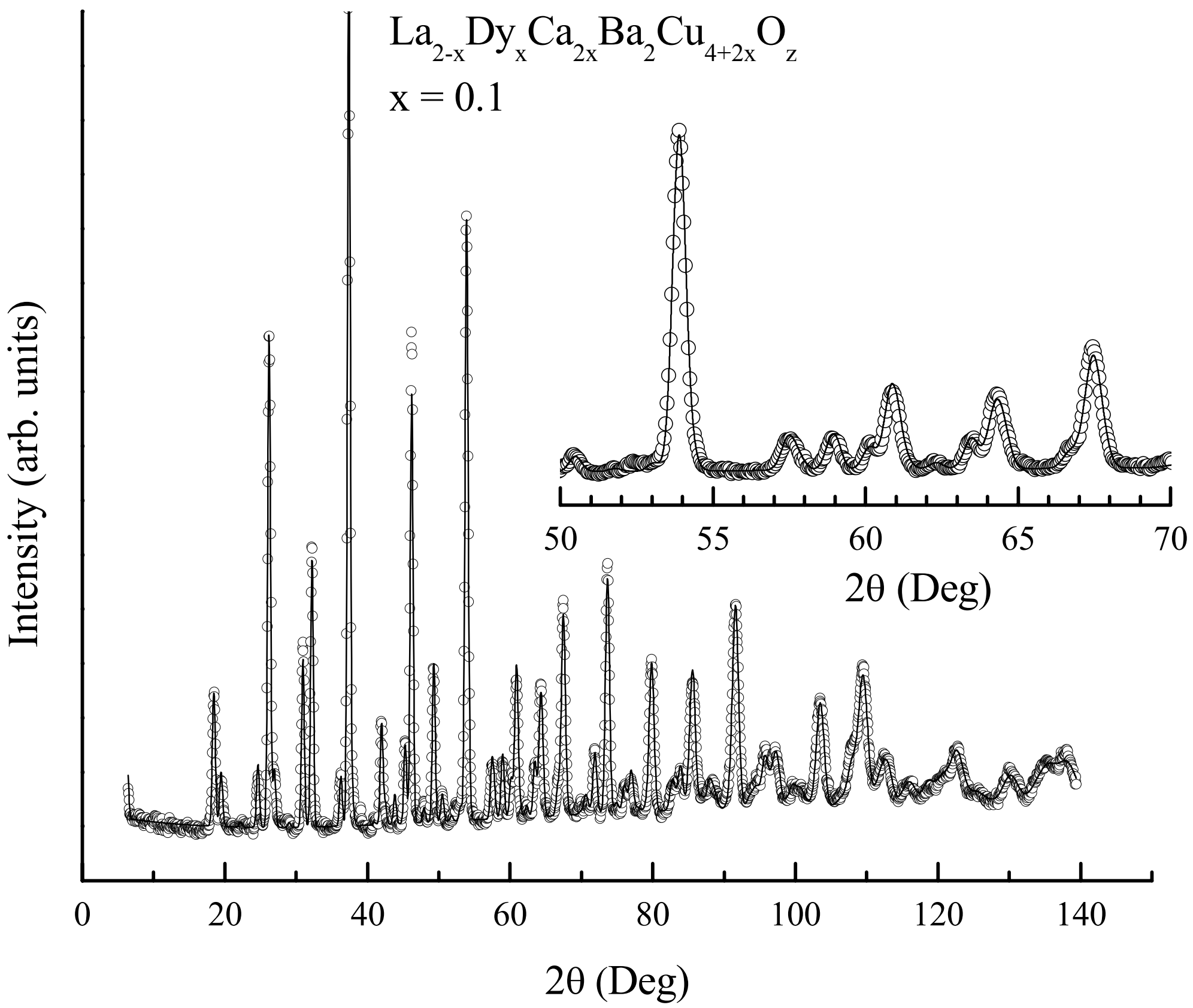

Figure 3

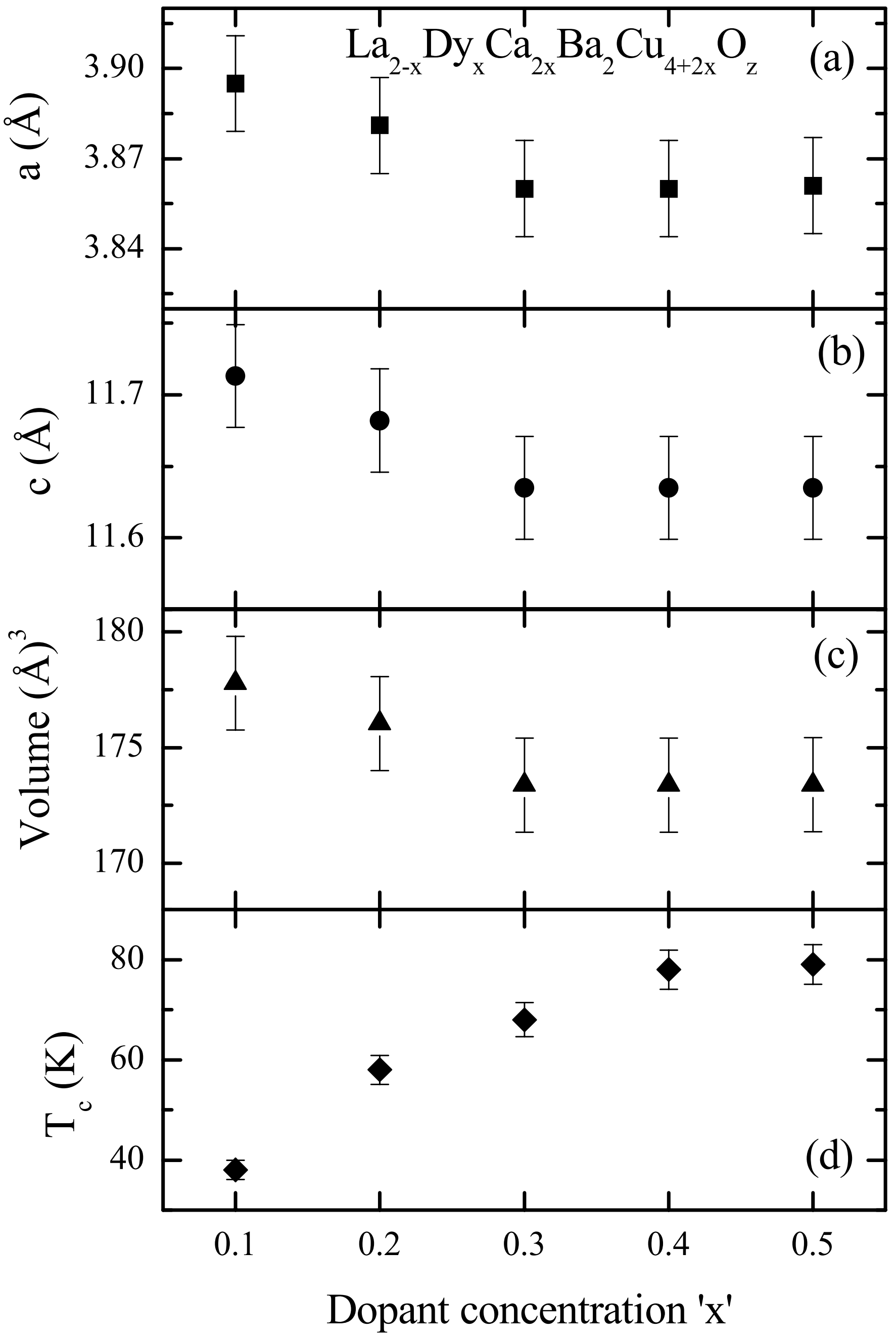